\documentclass[12pt]{article}

\usepackage{amssymb,amsmath} 
\usepackage{graphicx}
\usepackage{subfigure}

\usepackage{color}

\def\eeq{\relax}
\def\beq#1#2\eeq{\begin{equation}\label{#1}#2\end{equation}}
\def\bal#1#2\eal{\begin{align}\label{#1}#2\end{align}}
\def\bse#1#2\ese{\begin{subequations}\label{#1}#2\end{subequations}}

\def\bd#1{\mbox{\boldmath$\displaystyle\mathbf{#1}$} }
\def\div{\operatorname{div}} 
\def\grad{\operatorname{grad}} 
\def\Grad{\operatorname{Grad}} 
\def\Div{\operatorname{Div}} 
\def\det{\operatorname{det}} 
\def\dd{\operatorname{d}} 
\def\tr{\operatorname{tr}}

\def\ee{\mathrm{e}}
\def\cS{{\cal Q}}

\begin{document} 
\def\singlespacing{\baselineskip=13pt}	\def\doublespacing{\baselineskip=18pt}
\doublespacing

\title{Elastic cloaking theory}  

\author{A. N. Norris$^{a}$ and A. L. Shuvalov$^{b}$
\\ \\
$^{a}$ Mechanical and Aerospace Engineering, Rutgers University,\\
Piscataway, NJ 08854-8058, USA
\\
$^{b}\ $Universit\'{e} de Bordeaux, CNRS, UMR 5469, \\
Laboratoire de M\'{e}canique Physique, Talence 33405, France
}


\maketitle

\begin{abstract}   

Transformation theory is developed for the equations of linear anisotropic elasticity.    The transformed equations correspond to   non-unique material properties that can be varied for a given transformation by selection of the  matrix relating  displacements in the two descriptions. This gauge matrix can be chosen to make the   transformed density   isotropic for any transformation although the stress in the transformed material is not generally symmetric.   Symmetric stress is obtained only if the gauge matrix is identical to the transformation matrix, in agreement with Milton et al.
\cite{Milton06}.  
The elastic transformation theory is applied to the case of cylindrical anisotropy.   The equations of motion for the transformed  material with isotropic density are expressed in Stroh format, suitable for modeling cylindrical elastic cloaking.  It is shown that  there is a preferred approximate material with symmetric stress that could be a useful candidate for making cylindrical elastic cloaking devices. 

\end{abstract}

\newpage

\section{Introduction}\label{sec1}

Interest in cloaking of wave motion has   surged with the demonstration of the possibility of practical electromagnetic wave cloaking  \cite{Schurig06}.   The  principle underlying the effect  is the so-called transformation or change-of-variables method \cite{Greenleaf03,Pendry06} in which the material parameters in the physical domain are defined by a spatial transformation.  The concept of   material properties  defined by  transformation is not restricted to electromagnetism, and has stimulated interest in applying the same method to other wave fields.  The first applications in acoustics  were obtained by direct analogy with the electromagnetic case \cite{Cummer07,Chen07,Cummer08}.  It was quickly realized that the fundamental  mathematical identity behind the acoustic transformation is the  equivalence  \cite{Greenleaf07}  of the Laplacian in the original coordinates to a differential operator  in the transformed (physical) coordinates, according to 
$
 \Div\Grad f  \rightarrow   J\div J^{-1} {\pmb F}{\pmb F}^t\grad f$,
   where ${\pmb F} = \partial {\pmb x}/\partial {\pmb X}$ is the deformation gradient of the transformation (see \S \ref{sec2}) and $J=\det {\pmb F}$.  This connection, plus the realization that the tensor within the operator can be interpreted as a tensor of inertia  means that the homogeneous acoustic wave equation can be mapped to the equation for an inhomogeneous fluid with anisotropic density.   

However, the material properties for acoustic cloaking  do not have to be identified as a fluid with a single bulk modulus and a tensorial inertia.  There is a large degree of freedom in the choice of the cloaking material properties \cite{Norris08b,Norris09}: a  compressible fluid with anisotropic density is a special case of {\it pentamode materials} \cite{Milton95,Milton01} with anisotropic inertia.   The non-uniqueness of the material properties (for a given transformation) is a feature not found in the electromagnetic case, where, for instance, the tensors of electric permittivity  and  magnetic permeability  are necessarily proportionate for a transformation of the vacuum.  The extra freedom in the acoustic case means  that either or both of the scalar parameters, density and elastic stiffness (bulk modulus), can become tensorial quantities after the transformation.    While most papers on acoustic cloaking have  considered  materials with scalar stiffness and tensorial inertia, e.g.  \cite{Cummer07,Chen07,Cummer08,Cai07,Cummer08b,Cheng08,Norris08c,Pendry08a,Torrent08b}, see \cite{Chen10a} 
for a review, there is no physical reason for such restricted material properties.  Cloaking with such materials requires very large total mass \cite{Norris08b,Norris10a}, but the more general class of    pentamode materials  with anisotropic inertia does not have this constraint.  In fact, it is often possible to choose the material properties so that the inertia is isotropic, in which case the total mass is simply the mass of the equivalent undeformed region \cite{Norris09}. 
This property can be realized if the transformation is a pure stretch, as is the case when there is radial symmetry. 
This distinction between the cloaking material properties is critical but, judging by the continued emphasis on anisotropic inertia in the literature,  does not seem to have been fully  appreciated.  
Apart from  \cite{Norris08b,Norris09} there have been few studies \cite{Hu10,Scandrett10} of acoustic cloaking with anisotropic stiffness.

 The non-uniqueness of the transformed material properties  found in the acoustic theory transfers to elastodynamics.   The first study of the transformation of the elastodynamic equations by  Milton et al.  \cite{Milton06} concluded that the appropriate class of constitutive equations for the transformed material are the Willis equations  for material response.  
The general form of the Willis equations are \cite{Milton06,Milton07}
\bse{-272}
\bal{93-4}
 \div {\pmb \sigma} &= \dot{\pmb p}, 
\\
{\pmb \sigma} &= {\pmb C}^\text{eff} *{\pmb e}+{\pmb S}^\text{eff}*{\pmb u}, 
\label{93-4b}
\\
{\pmb p} &= {{\pmb S}^\text{eff}}^\dagger *{\pmb e} + {\pmb \rho}^\text{eff}*\dot {\pmb u},
\eal
\ese
where ${\pmb e} = \frac12( \nabla {\pmb u} +(\nabla {\pmb u})^t)$, $*$ denotes time convolution and $^\dagger$ is the adjoint. The  stress in \eqref{93-4b} is symmetric, and the  elastic moduli enjoy all of the symmetries for normal elasticity, \emph{viz.} 
$C_{klij}^\text{eff} = C_{ijkl}^\text{eff}$ and $C_{jikl}^\text{eff} = C_{ijkl}^\text{eff}$.  
Brun et al. \cite{Brun09} considered the transformation of isotropic elasticity in cylindrical coordinates for the  particular transformation function used by 
\cite{Greenleaf03,Pendry06} and  found that the transformed material properties are those of a material with  isotropic inertia and  elastic behavior of Cosserat type.  The governing equations for Cosserat elastic materials \cite{Cosserat} are 
\bse{=35}
\bal{354}
 \div {\pmb \sigma} &= \rho^\text{eff} \, \ddot{\pmb u}, 
\\
{\pmb \sigma} &= {\pmb C}^\text{eff}  \nabla{\pmb u},
\eal
\ese
with elastic moduli satisfying the major symmetry $C_{klij}^\text{eff} = C_{ijkl}^\text{eff}$ but not the minor symmetry,    $C_{jikl}^\text{eff} \ne C_{ijkl}^\text{eff}$.  This implies that the stress is not necessarily symmetric, ${\pmb \sigma}^t \ne {\pmb \sigma}$, and that it depends not only on the strain ${\pmb e}$ (the  symmetric part of the displacement gradient) but also upon the local rotation 
${\pmb \omega} = \frac12( \nabla {\pmb u} -(\nabla {\pmb u})^t)$.
Not only are the parameters such as ${\pmb C}^\text{eff}$ in eqs. \eqref{-272} and \eqref{=35} different, the constitutive theories are mutually incompatible: one has symmetric stress, the other a non-symmetric stress. 
We show in this paper that both theories are possible versions of the transformed elastodynamic equations, and that they are only two from  a spectrum  of possible constitutive theories.  
 Apart from the two references mentioned \cite{Milton06,Brun09}, the only other example of  transformation elasticity concerns flexural waves obeying the biharmonic equation \cite{Farhat09}, which  is beyond the realm of the present paper. 

The purpose here is to consider the transformation method for elastodynamics, and to describe the range of constitutive theories possible. 
 The starting point is the observation \cite{Norris09} that the extra degrees of freedom noted  for the acoustic transformation can be ascribed to  the linear relation between the displacement fields in the two coordinate systems.  
This "gauge" transformation introduces a second matrix or tensor, in addition to the deformation gradient from  the change of coordinates.  As discussed in \cite{Norris09}, the variety of acoustically transformed material properties arises from the freedom in  the displacement gauge.   
The same freedom is also present in the elastic case, and as we will show, it allows one to {derive a broader class of constitutive properties than those suggested by  Milton et al.  \cite{Milton06} and by Brun et al. \cite{Brun09}.  The material properties found in these studies correspond to specific choices of the gauge matrix.  }

 Cloaking is achieved with  transformations that deform  a region  in such a way that the mapping is one-to-one everywhere except at a single point, which is mapped into the cloak inner boundary.  This is a singular transformation, and in practice, the mapped region would be of finite size, e.g. a small sphere, for which the mapping is everywhere regular.  Our objective here is  to understand the nature of the material necessary to produce the transformation effect,  in particular, what type of constitutive behavior is necessary: such as isotropic or  anisotropic inertia.

The outline of the paper is as follows.   The notation and setup of the problem are 
given  in \S \ref{sec2} where the displacement gauge matrix is introduced.  The general form of the transformed equations of motion are presented in \S \ref{sec2.5}.  Constitutive equations resulting from specific forms of the gauge matrix are discussed in \S \ref{sec3}, particularly the Willis equations and Cosserat elasticity, which  are shown to coincide  under certain circumstances. 
The special case of transformed acoustic materials is discussed in \S \ref{sec4}. 
The elastic transformation theory with isotropic density is applied in  \S \ref{sec5} to radial transformation  of cylindrically anisotropic  solids.  Based on this formulation,  an elastic material with isotropic density and standard stress-strain relations is proposed in \S \ref{sec6}  as an approximation to the transformed material. 
 	A summary of the main results is given in 
\S \ref{sec7}.


\section{Notation and setup}\label{sec2}

Two related configurations are considered: the original $\Omega$, and the transformed region $\omega$, also called the physical or current domain. 
 The transformation from $\Omega$ to $\omega$ is described by the point-wise deformation 
from  ${\pmb X}\in \Omega$ to   ${\pmb x} \in \omega$. 
The symbols $\nabla$, $\nabla_X$ and $\div$, $\Div$  indicate the gradient and divergence operators in ${\pmb x}$  and ${\pmb X}$, respectively, and the superscript $t$ denotes transpose.  The component form of $\div {\bf E}$ is $\partial E_i /\partial x_i$ or 
 $\partial E_{ij} /\partial x_i$ when  ${\bf E}$ is a vector and a second order tensor-like quantity, respectively.    Upper and lower case subscripts $(I,J,\ldots , i,j,\ldots)$  are used to distinguish between the domains, and the summation convention on repeated subscripts is assumed.  It is useful to describe  the transformation using  language of finite 
deformation in continuum mechanics.  Thus, ${\pmb X}$ describes a particle position  in the Lagrangian or undeformed configuration, and 
${\pmb x}$ is particle location in the Eulerian or   deformed  physical state.  The transformation or mapping is assumed to be one-to-one and invertible.  For perfect cloaking the   transformation is one-to-many at the single point ${\pmb X}={\pmb O}$, but this can be avoided by always considering near-cloaks, where, for instance, the single point is replaced by a small hole which is mapped to a much larger hole.  
 
The deformation gradient is defined 
${\pmb F} = \nabla_X {\pmb x}$  with inverse ${\pmb F}^{-1} = \nabla {\pmb X}$, 
or in component form $F_{iI}  = \partial x_i /\partial X_I$, $F_{Ii}^{-1}  = \partial X_I /\partial x_i$.    
The Jacobian of the deformation is $J = \det {\pmb F} = |{\pmb F}|$,  or in terms of volume elements in the two 
configurations, $J = \dd v/\dd V$. The  polar decomposition implies ${\pmb F} = {\pmb V} {\pmb R} $, where ${\pmb R}$
 is proper  orthogonal ($ {\pmb R}{\pmb R}^t = {\pmb R}^t {\pmb R} = {\pmb I}$, $\det {\pmb R}= 1$) and the left 
stretch tensor ${\pmb V}\in$ Sym$^+$ is the  positive definite solution of 
${\pmb V}^2 = {\pmb B}$ where ${\pmb B}$ $(= {\pmb F}{\pmb F}^t)$ 
is the left Cauchy-Green or Finger deformation tensor.

The  infinitesimal  displacement  ${\pmb U}({\pmb X},t) $ and   stress  
${\pmb \Sigma}({\pmb X},t) $ 
satisfy the equations of linear elasticity  in the original domain: 
\beq{91}
\left.
\begin{split}
\Div {\pmb \Sigma}  &=  \rho_0\ddot{\pmb U} ,
\\
 {\pmb \Sigma} &= {\pmb C}^{(0)} \nabla_X {\pmb U},
\end{split}
\right\} \quad \text{in } \Omega ,
\eeq
where $\rho_0$ is the  (scalar) mass density 
 %
and the 
 the elements of the elastic stiffness tensor satisfy the full symmetries
\beq{2}
C^{(0)}_{ IJKL}=C^{(0)}_{ JIKL}, \quad C^{(0)}_{IJKL}=C^{(0)}_{KLIJ}.
\eeq
The first identity expresses the symmetry of the stress 
and the second is the consequence of an assumed strain energy density function. 
The total energy density is the sum of the strain and kinetic energy densities, 
 \beq{4}
\mathcal{E}_0 = \mathcal{W}_0+\mathcal{T}_0\quad \text{where  }
\mathcal{W}_0 = \tfrac 12 C^{(0)}_{ IJKL} \, U_{J,I}U_{L,K}, 
\quad \mathcal{T}_0 = \tfrac 12 \rho_0 \dot{\pmb U}^t\dot{\pmb U}. 
\eeq

Particle displacement in the transformed domain is ${\pmb u}({\pmb x},t)$.  Our objective is to find its governing equations.  In order to proceed, we need in addition to the geometrical quantity ${\pmb F}$, a kinematic relation that relates the displacements in the two domains. We assume a linear 
"gauge" change in the displacement defined by a non-singular matrix ${\pmb A}$ as 
\beq{2.1}
{\pmb U} = {\pmb A}^t {\pmb u} \qquad (U_I=A_{iI}u_i).
\eeq
According to  its definition the matrix ${\pmb A}$ is, like ${\pmb F}$, not a second order tensor because it has one "leg'' in both domains.  The choice of the transpose, ${\pmb A}^t$ in equation \eqref{2.1}, means that   ${\pmb A}$ and ${\pmb F}$  are similar objects, although at this stage they are not related. 

The arbitrariness in the choice of ${\pmb A}$  is the central theme  of this paper.  
This approach generalizes that of 
\cite{Norris09} which was restricted to   acoustic materials, and of  Milton et al.  \cite{Milton06} for elasticity. The point of departure with \cite{Milton06} here rests with the assumed independence of the  gauge matrix ${\pmb A}$.   Milton et al.   assume a similar relation  between the displacement fields; eq. \eqref{2.1} is identical to eq. (2.2) in \cite{Milton06}; however, the matrix ${\pmb A}$ in \cite{Milton06} is assumed at the outset to be equal to the deformation gradient   $({\pmb A}={\pmb F})$.  We will return to this distinction later.  As noted by  
Milton et al.  \cite{Milton06}, the relation 
$
\dd {\pmb X } = {\pmb F}^{-1} \dd {\pmb x } 
$
might lead one to expect ${\pmb A}={\pmb F}^{-t}$ by identifying 
$\dd {\pmb X }$ and $\dd {\pmb x }$ with ${\pmb U }$ and ${\pmb u }$, respectively.  However, the displacements are not associated with the coordinate transformation, unlike in the theory of finite deformation, and hence $ {\pmb F}$ and $ {\pmb A}$ are independent.  Milton et al.  \cite{Milton06}  specify ${\pmb A}={\pmb F}$ on the basis that this is the only choice that guarantees a symmetric stress.  We will return to this point later. 

\section{General form of the transformed equations}\label{sec2.5}
Under the transformation and the gauge change the energy density transforms as $\mathcal{E}_0 \rightarrow \mathcal{E} = \mathcal{W}+\mathcal{T}$ according to 
$
\mathcal{E}\dd V = \mathcal{E}_0 \dd V_0$, 
so that
\beq{6}
\mathcal{E}=\mathcal{W}+\mathcal{T} = \tfrac 12 
J^{-1} \big\{
C^{(0)}_{ IJKL}\, \big(u_j A_{jJ}\big)_{,i}\big(u_l A_{lL}\big)_{,k}
F_{iI}F_{kK} 
+ \rho_0 \dot{u}_i \dot{u}_j A_{iI}A_{jI}.
\big\}
\eeq
Hence, 
\beq{7}
\mathcal{W}= \tfrac 12 J^{-1} C^{(0)}_{ IJKL}\, F_{iI}F_{kK} \, 
\big(u_j A_{jJ}\big)_{,i}\big(u_l A_{lL}\big)_{,k}
 ,
\qquad
\mathcal{T} = \tfrac 12 \dot{\pmb u}^t {\pmb \rho}\dot{\pmb u},  
\eeq
where the (symmetric) density tensor is 
\beq{8}
{\pmb \rho} = {\pmb \rho}^t = \rho_0 \, J^{-1}  {\pmb A}{\pmb A}^t ,
\eeq
The equations of motion in the  
deformed, or current material, are determined by the Euler-Lagrange equations of the Lagrangian density 
$\mathcal{L}=\mathcal{W}-\mathcal{T}$, as 
\beq{9.12}
 A_{jJ}\, \big(J^{-1} C^{(0)}_{ IJKL}\, F_{iI} (u_l A_{lL})_{,k} F_{kK} \big)_{,i} - \rho_{ij}\ddot{u}_i=0 .
\eeq
Using the identity 
\beq{0-2}
(J^{-1}  F_{iI}  )_{,i} =0,
\eeq
 this can be written 
\beq{9.1}
 \cS_{ijIJ}  \, \big(J C^{(0)}_{ IJKL}\,(u_l \cS_{klKL})_{,k}  \big)_{,i} - \rho_{ij}
\ddot{u}_i=0 ,
\eeq
where the fourth order quantity
\beq{2-1}
\cS_{ijIJ} 
= J^{-1} F_{iI} A_{jJ}, 
\eeq
is introduced for later use.

{The transformed system \eqref{9.1} is equivalent  to the equilibrium equations
\bse{4-4}
\beq{-5=}
\sigma_{ij,i} = \dot{p}_j,
\eeq
and  the  constitutive relations 
\beq{-50}
\sigma_{ij}=C_{ijkl}^\text{eff}u_{l,k}+S_{ijl}^\text{eff}\, \dot{u}_l,
\quad
p_l = S_{ijl}^\text{eff}\, u_{j,i} + \rho_{jl}^\text{eff}\dot{u}_k, 
\eeq
\ese
with parameters defined as follows in the Fourier time domain (dependence $e^{-i\omega t}$)
\bse{-51}
\bal{-51a}
C_{ijkl}^\text{eff} &
= J C^{(0)}_{ IJKL} \cS_{ijIJ} \cS_{klKL},
\\
S_{ijl}^\text{eff} &
= (-i\omega )^{-1} J C^{(0)}_{ IJKL} \cS_{ijIJ} \cS_{klKL,k},
\\
\rho_{jl}^\text{eff}&
= \rho_{jl} + (-i\omega )^{-2}  J C^{(0)}_{ IJKL} \cS_{ijIJ,i} \cS_{klKL,k}
, 
\eal
\ese
where the  density $\rho_{jl}$ is given by \eqref{8}. 
The elastic moduli and the density  satisfy the general symmetries 
\beq{03-}
C_{ijkl}^\text{eff}  = C_{klij}^\text{eff} ,
\quad
\rho_{jl}^\text{eff} = \rho_{lj}^\text{eff}, 
\eeq
but not the full symmetries required for the Willis constitutive model \eqref{-272}.  
Equations \eqref{4-4}-\eqref{-51} are the fundamental result of the transformation theory.   The remainder of the paper is concerned with their simplification and interpretation. 
}

Note that the transformed stiffness may be expressed in a form similar to  
the Kelvin representation for the tensor of elastic moduli \cite{kelvin}, as
\beq{2-7}
{\bd C}^\text{eff} =  \sum_{\alpha = 1}^6 K^{(\alpha)}
{\bd S}^{(\alpha)}  \otimes {\bd S}^{(\alpha) },
\quad 
K^{(\alpha)} = J K_0^{(\alpha)},
\quad 
{\bd S}^{(\alpha)}  = 
J^{-1} {\bd F}
{\bd P}^{(\alpha)} {\bd A}^t ,
\eeq
where 
 $K_0^{(\alpha)} >0$ are the Kelvin moduli, ${\bd P}^{(\alpha) } \in$ Sym,  $\tr 
{\bd P}^{(\alpha) } {\bd P}^{(\beta) } =\delta_{\alpha\beta}$,  are the eigenstrains/eigenstresses, such that the original stiffness  has the unique decomposition
\beq{2-6}
{\bd C}^{(0)} =  \sum_{\alpha = 1}^6 K_0^{(\alpha)}
{\bd P}^{(\alpha)}  \otimes {\bd P}^{(\alpha) } . 
\eeq  
The transformed matrices ${\bd S}^{(\alpha)}$   do not inherit the orthogonality of the original eigenstrains/eigenstresses 
 ${\bd P}^{(\alpha) }$, so that \eqref{2-7} is not the exact Kelvin representation in the transformed coordinates.  It does, however, illustrate that the transformed stiffness is positive definite, even though ${\bd S}^{(\alpha)}$ 
 are in general not symmetric.    The representation \eqref{2-7} is particularly useful in the limiting case  of an acoustic fluid in the original domain for which only one of the $K_0^{(\alpha)}$ is non-zero, discussed later.

{
\section{Transformed equations in specific forms}\label{sec3}
\subsection{Willis equations: ${\pmb A} =   {\pmb F}$ }  \label{sec4.1}
The absence of the minor symmetries under the interchange of $i$ and $j$ in 
$C_{ijkl}^\text{eff}$ and $S_{ijl}^\text{eff}$ of \eqref{-51} implies that the stress is generally not symmetric.  Symmetric stress is guaranteed if $\cS_{ijIJ} = \cS_{jiIJ}$ (see eq. \eqref{2-1}), which occurs if the gauge matrix is of the form $
{\pmb A} = \zeta {\pmb F} $,
for any scalar $\zeta \ne 0$, which may be set to unity with no loss in generality.  This recovers the results of Milton et al. \cite{Milton06} that the   transformed material is of   the Willis form, eq. \eqref{-272}.   As  noted in \cite{Milton06}, this is the only choice for ${\pmb A}$ that yields symmetric stress.   
}

In summary, the governing equations are \eqref{4-4}  with material parameters defined by  \eqref{-51} and 
\beq{8-1}
\cS_{ijIJ} 
= J^{-1} F_{iI} F_{jJ}.  
\eeq
The  parameters now display  the full symmetries expected of  a Willis material: 
\beq{03-8}
C_{ijkl}^\text{eff}  = C_{klij}^\text{eff} ,
\quad
C_{ijkl}^\text{eff}  = C_{jikl}^\text{eff} ,
\quad
\rho_{jl}^\text{eff} = \rho_{lj}^\text{eff}
\quad
S_{ijl}^\text{eff}  = S_{jil}^\text{eff} .
\quad.
\eeq
Note that the  stiffness tensor is 
\bal{94}
C_{ijkl}^\text{eff}
&= F_{iI}F_{jJ} F_{kK}F_{lL}\, J^{-1}  C^{(0)}_{ IJKL} 
= V_{iI}V_{jJ} V_{kK}V_{lL}\, J^{-1}  \bar{C}^{(0)}_{ IJKL} , \text{  where  }
\nonumber  \\
\bar{C}^{(0)}_{ IJKL}   &= R_{IM}R_{JN}R_{PK}R_{LQ}\, C^{(0)}_{ MNPQ}
\eal
are the original moduli in the rotated frame.   The full symmetry of the stiffness tensor also follows immediately from the  representation \eqref{2-7} with symmetric 
${\bd S}^{(\alpha)}  = 
J^{-1} {\bd F} {\bd P}^{(\alpha)} {\bd F}^t$.

\subsection{Cosserat elasticity: ${\pmb A} = $ constant}
\subsubsection{General form}
The constitutive parameters \eqref{-51} simplify considerably if the fourth order quantity $\pmb \cS$ satisfies 
\beq{2-2}
\cS_{ijIJ,i} = 0.
\eeq
This differential constraint combined with  \eqref{0-2} implies that the gauge ${\bd A}$ {\it must be constant}.  In that case the transformed equations of motion become
\beq{12-8}
\sigma_{ij,i}  =  \rho_{ij}\, \ddot{u}_j, 
\quad 
\sigma_{ij} = C_{ijkl}^\text{eff}\, u_{l,k},
\eeq
where the effective elastic moduli are defined by \eqref{2-1} and
\eqref{-51a} 
and the density tensor $\pmb \rho$ is given in \eqref{8}.

Note that the elastic moduli satisfy the symmetry \eqref{03-}$_1$ 
associated with the transformed energy density 
$\mathcal{W}= \tfrac 12 C_{ijkl}^\text{eff} u_{j,i}u_{l,k}$. 
But $C_{ijkl}^\text{eff}$   does not satisfy the minor symmetry \eqref{2}$_1$ since 
\bal{14}
C_{ijkl}^\text{eff}-C_{jikl}^\text{eff} &= J^{-1} C^{(0)}_{ IJKL}\,F_{kK}A_{lL}  \big( F_{iI}A_{jJ} -F_{jI}A_{iJ} \big)
\nonumber \\
&= J^{-1} C^{(0)}_{ IJKL}\,F_{kK}A_{lL}  \big( F_{iI}A_{jJ} -F_{jJ}A_{iI} \big), 
\eal
which is non-zero in general (note that the second form in \eqref{14} uses the minor symmetry \eqref{2}$_1$ for the original moduli $C^{(0)}_{ IJKL}$).  This means that the stress is not necessarily symmetric, $
{\pmb \sigma}\ne {\pmb \sigma}^t$,
 which places the material in the framework of Cosserat elasticity \cite{Cosserat}.  The number of independent elastic stiffness elements is at most $9(9+1)/2=45$ as compared with $6(6+1)/2 = 21$ for normal linear elasticity. 

\subsubsection{Cosserat elasticity with isotropic density: ${\pmb A} =   {\pmb I}$   }
Isotropic density can  be achieved by taking the constant matrix  ${\pmb A}$   proportional to the identity, ${\pmb A}=\zeta {\pmb I}$, with $\zeta =1$ without loss of generality.  In this important case we have 
${\pmb \rho} = \rho {\pmb I} $, with 
\beq{81}
 \rho = \rho_0 \, J^{-1}  ,
\quad
 C_{ijkl}^\text{eff} 
=J^{-1} C^{(0)}_{ IjKl}\,  F_{iI} F_{kK} .  
\eeq

\subsection{Examples}

\subsubsection{Example 1: SH motion in a plane of material symmetry}\label{-0-0}
The original moduli are assumed to have a plane of symmetry perpendicular to the $X_3$-axis, and the transformation  is assumed to preserve the out of plane coordinate: $x_3 = X_3$.   Consider shear horizontal motion ${\pmb U} = (0,0,U(X_1,X_2,t))$ satisfying the scalar equation
\beq{033=}
\big( C^{(0)}_{A3B3} U_{,B}
\big)_{,A} = \rho_0 \ddot{U}, 
\eeq
with  indices $A,B\in\{1,2\}$.  Under these circumstances, the  equation for SH motion in the transformed domain, ${\pmb u} = (0,0,u(x_1,x_2,t))$, is the same for  both the Willis constitutive equations $({\pmb A} = {\pmb F})$ and the Cosserat model with isotropic density $({\pmb A} = {\pmb I})$.    Thus, 
\beq{034=}
\big( C^\text{eff} _{\alpha 3\beta 3} u_{,\beta}
\big)_{,\alpha} = \rho \ddot{u}, 
\eeq
where  $\alpha ,\beta \in\{1,2\}$,  $\rho = J^{-1}\rho_0$, and 
$C^{\text{eff}}_{\alpha 3\beta 3} = J^{-1} C^{(0)}_{A3B3} F_{\alpha A}F_{\beta B}$.  
The equivalence may be expected since the only relevant element of the gauge matrix, $A_{33}$, is the same  for both models  $(A_{33} = 1)$. 

The above conclusion for SH motion in the presence of  orthotropic moduli relies only upon the scalar nature of the motion in the original and transformed domains.  As such, the 
SH results also  follow from those in \S\ref{sec4} for fluid  acoustics
under the standard replacement of fluid  density and bulk modulus with inverse
shear modulus and inverse solid density, respectively.

\subsubsection{Example 2:  ${\pmb F} = $ constant}

When both  ${\pmb A}$ and  ${\pmb F}$ are constant the Willis equations simplify to  those of normal linear elasticity $(S_{ijl}=0)$. 
 At the same time, the constant deformation gradient implies that   $ {\pmb A}=    {\pmb F}$ (=constant) is a permissible choice for the Cosserat medium.  The Willis and Cosserat materials are then coincident with 
 density 
\beq{813}
{\pmb \rho} = {\pmb \rho}^t = \rho_0 J^{-1} \,  {\pmb B}.
\eeq
and fully symmetric elastic moduli given by \eqref{94}. 
Note that the density ${\pmb \rho}$ is  anisotropic unless ${\pmb B} = \alpha {\pmb I}$, which means the deformation is a pure expansion, possibly with rotation.  But this is a rather trivial case.

Consider an isotropic initial material  with original moduli  $C^{(0)}_{ IJKL}  = \lambda_0 \delta_{IJ}\delta_{KL}+\mu_0(\delta_{IK}\delta_{JL}+\delta_{IL}\delta_{JK})$, 
or equivalently, 
\beq{32}
{\pmb C}^{(0)} = \lambda_0 \, {\pmb I}\otimes{\pmb I}  +2\mu_0 \,  {\pmb I}\boxtimes{\pmb I} , 
\quad \text{where  }
\big( {\pmb X}\boxtimes{\pmb X}\big)  {\pmb Y}\equiv \frac12  {\pmb X}\big( {\pmb Y}+{\pmb Y}^t\big) {\pmb X}. 
\eeq
The rotated moduli of \eqref{94}$_2$ are therefore unchanged, $\bar{\pmb C}^{(0)} = {\pmb C}^{(0)}$, and the current density and  moduli  are 
\beq{33}
{\pmb C}  = \lambda  \, {\pmb B}\otimes{\pmb B}  +2\mu  \,  {\pmb B}\boxtimes{\pmb B} , 
\quad
{\pmb \rho} =  \rho \, {\pmb B},
\quad
\text{ where  }\{\lambda  , \mu  ,\rho  \}=  J^{-1} \, \{\lambda_0  , \mu_0  ,\rho_0  \}.
\eeq 
It is of interest to consider  plane wave motion, $
{\pmb u} = {\pmb q}\, g({\pmb n}^t {\pmb x} - v t)$ 
 for unit vector ${\pmb n}$,  constant ${\pmb q} $, and $g\in \mathrm{C}^2$.
The equation of motion \eqref{12-8}$_1$ implies that the polarization vector ${\pmb q}$ satisfies   
\beq{36}
\big( {\pmb Q}  - v^2 {\pmb \rho} \big)\,  {\pmb q}=0, 
\quad
\text{ where  }{\pmb Q} = (\lambda+\mu) ( {\pmb B}{\pmb n})\otimes( {\pmb B}{\pmb n}) 
+\mu ({\pmb n}^t {\pmb B}{\pmb n})  \, {\pmb B} .
\eeq
The solutions of \eqref{36} are of two distinct types:
\bse{37}
\bal{37b}
\text{longitudinal:}\quad &{\pmb q}\parallel {\pmb n},\quad v= ({\pmb n}^t {\pmb B}{\pmb n})^{1/2}\, c_L, 
\quad c_L^2 = \frac{\lambda+2\mu}{\rho}=   \frac{\lambda_0+2\mu_0}{\rho_0},   
\\
\text{quasi-transverse:}\quad &{\pmb q}\perp{\pmb B}{\pmb n},\quad v= ({\pmb n}^t {\pmb B}{\pmb n})^{1/2}\, c_T, 
\quad c_T^2 = \frac{ \mu}{\rho}=  \frac{ \mu_0}{\rho_0}. 
\eal
\ese
The slowness vector is ${\pmb s} = v^{-1}{\pmb n}$, and the slowness surface is the envelope of the slowness vectors for all propagation directions ${\pmb n}$.  The slowness surface is therefore comprised of two sheets which are similar ellipsoids: ${\pmb s}^t {\pmb B}{\pmb s} = c_\alpha^{-2}$, $\alpha=L,T$. Note that the polarizations do not in general form an orthogonal triad. 

\section{Acoustics as a special case}\label{sec4}

\subsection{General form of transformed equations}

We consider the simpler but special case of acoustics in order to further understand the structure of the  elastic transformation theory.   The acoustic equations are unique in the sense that they are the single  example of a pentamode material  \cite{Milton95,Milton01} commonly encountered  in mechanics.  We will demonstrate that the pentamode property introduces a unique degree of freedom not available in the fully elastic situation.  

The elastic stiffness of an acoustic fluid is 
\beq{--3}
{\bd C}_0 =K_0 \, {\bd I}\otimes {\bd I}
, 
\eeq
which is of pentamode form \cite{Milton95}, i.e. the 6$\times$6 Voigt matrix associated with the elements 
$C^{(0)}_{ IJKL}= K_0 \delta_{IJ}\delta_{KL}$ has five zero eigenvalues. 
Equation  \eqref{9.1} becomes, using \eqref{--3}, 
\beq{9.14}
  Q_{ij}\, \big(J K_0 \,(u_l Q_{kl} )_{,k}  \big)_{,i} - \rho_{ij}\ddot{u}_i=0 ,
\eeq
with the density tensor given by \eqref{8}, and 
\beq{914}
{\bd Q}  = J^{-1} {\bd F}{\bd A}^t 
\quad \big( Q_{ij} =  J^{-1} F_{iI} A_{jI}  = \cS_{ijKK} 
\big) .
\eeq
The general form of the governing equations are again of the form \eqref{4-4}, with material parameters
\bse{-44}
\bal{-44a}
C_{ijkl}^\text{eff} &
= K_0 J^{-1} B_{ik} A_{jN} A_{lN},
\\
S_{ijl}^\text{eff} &
= (-i\omega )^{-1}  K_0 J^{-1} B_{ik} A_{jN} A_{lN,k},
\\
\rho_{jl}^\text{eff}&
= \rho_{jl} + (-i\omega )^{-2}   K_0 J^{-1} B_{ik} A_{jN,i} A_{lN,k} , 
\eal
\ese
recalling that  ${\pmb B}= {\pmb F}{\pmb F}^t$.   The stress is again not generally symmetric, unless  
${\pmb A}$ is proportional to ${\pmb F}$, in which case the transformed material is of Willis form, 
see \S\ref{sec4.1}. 

\subsection{Cosserat and related forms of the transformed equations}
The simplified form of the elasticity in the acoustic fluid implies that the condition \eqref{2-2}, which is required to simplify the form of the transformed equations, itself simplifies to the condition  that $\bd Q$ be divergence free:
\beq{333}
 \div{\bd Q}=0
\quad \big(  Q_{ij,i} = 0 \big) .
\eeq
Note that this is a necessary but not sufficient condition for the more general elasticity version \eqref{2-2}. 
Assuming that \eqref{333} holds, the transformed equations \eqref{9.14}  have the simplified Cosserat structure   \eqref{12-8}, with 
\beq{582}
{\bd \rho} =\rho_0 J \,   {\bd Q}^t {\bd B}^{-1} {\bd Q}, 
\quad  {\bd C}^\text{eff} =K_0 J \, {\bd Q}\otimes {\bd Q}, 
\quad  \div  {\bd Q} =0. 
\eeq 


We note the symmetries
${\pmb \rho} = {\pmb \rho}^t$,
$
C_{ijkl}^\text{eff}  = C_{klij}^\text{eff}$,
but the minor symmetry $C_{ijkl}^\text{eff}  = C_{jikl}^\text{eff}$ does not in general hold unless $\bd Q$ is symmetric \cite{Norris09}.  Thus, the  transformed acoustic equations are  those of  a pentamode material of Cosserat type with anisotropic density.  All previous studies of  transformation acoustics assumed {\it a priori} that the transformed materials must have symmetric stress.  The present results show that the more general structure of the transformed  properties is that of a   material with 
 stress  not necessarily  symmetric.   The pentamode structure of ${\bd C}^\text{eff}$ implies that the equations of motion    \eqref{12-8} can    be expressed in a form that is clearly related to acoustics \cite{Norris08b}  by using a scalar "pseudo-pressure" $p$ and "bulk modulus" $K = K_0 J$,  
\beq{-44=}
{\bd \rho} \dot{\bd v} = - {\bd Q}\nabla p, 
\quad
\dot{p} =- K \tr( {\bd Q}\nabla {\bd v}). 
\eeq

The condition \eqref{333} can be achieved, as in the elastic case, with constant gauge $\pmb A$.   For instance, taking  ${\pmb A} =  {\pmb I}$, yields material  properties  (see \eqref{81})  
\beq{322}
\rho = \rho_0 J^{-1}, 
\quad 
{\bd C}^\text{eff} =K_0 J^{-1} \, {\bd F}\otimes {\bd F}.
\eeq
This describes a material with isotropic density of general pentamode/Cosserat form. That is, the stiffness is pentamode (a single nonzero  eigenstiffness \cite{kelvin}) 
and a single eigenstress of generally non-symmetric form, hence Cosserat.     As discussed in \cite{Norris08b}, isotropic density with symmetric stress can be achieved if 
$\div h {\bd V} = 0$ for some function $h({\bd x})$.  One important case is when the deformation is a pure stretch ${\bd F} = {\bd V}$ (with $h=J^{-1}$, see eq. \eqref{0-2}), for which the transformed material   is pure pentamode  with isotropic density: 
\beq{902}
{\bd \rho} =\rho_0 J^{-1} \,   {\bd I}  , 
\quad  {\bd C} =K_0 J^{-1} \, {\bd V}\otimes {\bd V},
\quad \text{pure stretch}.
\eeq
 More  general conditions under which pure pentamode material with isotropic density can be achieved are discussed in 
\cite{Norris09}. 

Condition \eqref{333} may also be satisfied by non-constant gauge matrices.  For instance, 
${\bd A}  = J {\bd F}^{-t}$ gives ${\bd Q}  =   {\bd I}$ which clearly satisfies \eqref{333}. In this case the  transformed medium has the properties 
\beq{556}
{\bd \rho} =\rho_0 J \,   {\bd B}^{-1}  , 
\quad  {\bd C}^\text{eff} =K_0 J  \, {\bd I}\otimes {\bd I}.
\eeq
This corresponds to a fluid with isotropic (hydrostatic) stress, bulk modulus $K=K_0 J $, and anisotropic density $\bd \rho$.   This type of material was the first to be considered for acoustic cloaking 
\cite{Cummer07,Chen07,Cummer08}
yet it is remarkable that it does not possess a generalization to elasticity.  Rather, it is made possible by the simple structure of the second order quantity ${\pmb Q}$ as compared with its  fourth order elasticity analog  ${\pmb \cS}$.  The  condition \eqref{2-2} for ${\pmb \cS}$ is only satisfied by constant ${\bd A}$, but the analogous condition \eqref{333} for ${\pmb Q}$ has at least one non-constant solution for ${\pmb A}$.  The one noted here,  ${\bd A}  = J {\bd F}^{-t}$, yields the class of acoustic cloaking materials that use anisotropic inertia as the active mechanism. 

The transformed acoustic material can also be understood  as the special case in which five of the Kelvin moduli  $ K_0^{(\alpha)}$ vanish (see \eqref{2-6}), 
say $\alpha  =2,3,\ldots , 6$, 
and the remaining single eigenvector is the identity, ${\bd P}^{( 1) } = {\bd I}$.  The transformed material  has a single non-zero eigen-stiffnesses with associated eigen-tensor  (see eq. \eqref{914}) 
\beq{210}
{\bd S}^{(1)}  = 
J^{-1} {\bd F} {\bd A}^t.
\eeq
The inertial transformed material \eqref{556} results from the choice of ${\bd A}$ that makes 
${\bd S}^{(1)}  = {\bd I}$, that is 
${\bd A} = J {\bd F}^{-t}$.

\section{Applications in cylindrical elasticity}\label{sec5}

The non-uniqueness in the form of the transformed equations of elasticity provides the designer of elastic cloaking devices with a wide variety  of possible materials from which to choose.  These range from materials with Willis constitutive behavior \eqref{-272},  to Cosserat materials \eqref{81}, each special cases of the general constitutive relations  \eqref{4-4}-\eqref{-51}.  The same non-uniqueness exist for acoustic cloaking, where designs based on inertial cloaking on the one hand, and pentamode materials on the other, reflect the choice of completely different material properties.   One aspect common to all of these materials is their exotic nature.  In the absence of available materials with exactly the right properties it is reasonable to ask what can be achieved using "normal" materials, i.e. those with isotropic density and standard linear elastic response, including a symmetric stress.  

With that goal in mind we  examine in this section cloaking in the context of the Cosserat materials described by \eqref{81}.  This class of metamaterials is chosen as the starting point because of its property of isotropic density and the fact that the constitutive behavior is local.  That is, the stress depends on the displacement gradient alone, and not on displacement as in \eqref{4-4}-\eqref{-51}.  The goal is to find if there is a material with symmetric stress that provides an optimal, in some sense, approximation. We consider  transformations in cylindrical coordinates, starting with the formulation of the Cosserat constitutive equations in cylindrical coordinates.   Thus, for the remainder of the paper ${\bd A}$ is assumed to be a constant, and we use ${\pmb C}$ instead of $  {\pmb C}^\text{eff}$.  

\subsection{General theory for Cosserat materials}
%

\subsubsection{Cosserat notation in cylindrical coordinates}

The cylindrical coordinates are referred to by the  indices $1,2,3$ for $r, \theta, z$, respectively.   
The usual  Voigt notation, which  means $\{1,2,3,4,5,6\} = \{11,22,33,23,31,12\}$, 
 is augmented with three  
additional  indices to describe Cosserat elasticity: $\{\bar 4,\bar5,\bar 6\} = \{32,13,21\}$, so that the elastic stiffness tensor becomes in the 9-index Voigt-Cosserat notation, 
\beq{433}
{\pmb C} =\begin{pmatrix}
	c_{11} &c_{12} &c_{13} &c_{14}  &c_{1\bar 4}&c_{15}&c_{1\bar 5}&c_{16}&c_{1\bar 6} \\
	  &c_{22} &c_{23} &c_{24}  &c_{2\bar 4}&c_{25}&c_{2\bar 5}&c_{26}&c_{2\bar 6} \\
	  &  &c_{33} 
		&c_{34}  &c_{3\bar 4}&c_{35}&c_{3\bar 5}&c_{36}&c_{3\bar 6} \\
	       &       &      &c_{44}  &c_{4\bar 4}&c_{45}&c_{4\bar 5}&c_{46}&c_{4\bar 6} \\
         &       &       & &c_{\bar 4\bar 4}&c_{\bar 45}&c_{\bar 4\bar 5}&c_{\bar 4 6}&c_{\bar 4\bar 6} \\
         &      &      &       &           &c_{55}   &c_{5\bar 5}&c_{56}&c_{5\bar 6} \\
         &       &      &      &      &    &c_{\bar 5\bar 5}&c_{\bar 5 6}&c_{\bar 5\bar 6} \\
	      &   S   &   Y   &    M   &          &       &          &c_{66}   &c_{6 \bar  6}   \\
	      &      &      &      &          &        &        &    &c_{ \bar  6 \bar  6} 
\end{pmatrix}
\quad
\Leftrightarrow \quad
\begin{pmatrix}
1\\ 2\\ 3\\ 4 \\ \bar{4}\\ 5\\ \bar{5}\\ 6 \\ \bar{6}
\end{pmatrix}.
\eeq

Following \cite{Norris10}, the traction vectors 
 $\mathbf{t}_i=\mathbf{t}_i ({\mathbf x}, t)$, $i=r,\theta,z$,  are defined by the orthonormal basis vectors 
 $\{ \mathbf{e}_r, \mathbf{e}_\theta, \mathbf{e}_z\}$ 
 of the cylindrical coordinates 
 $\{r, \theta ,z\}$ according to 
 $\mathbf{t}_i   = \mathbf{e}_i \pmb{\sigma}$ $(i=r,\theta,z)$,  
 where $  \pmb{\sigma} ({\mathbf x} ,t)$ is the stress, 
and a comma denotes partial differentiation. 
 With  the same basis vectors,  and assuming the summation convention on repeated indices,  the 
elements of stress are $\sigma_{ij} = C_{ijkl} \varepsilon_{kl}$ where 
$\pmb{\varepsilon} = \frac12 (\nabla {\mathbf u} + \nabla {\mathbf u}^t)$ is the strain, 
$ C_{ijkl} = C_{ijkl} ({\mathbf x})$ are elements of the fourth order  (anisotropic) elastic stiffness tensor.    The traction vectors are    
\cite{Shuvalov03nonote}
{\small 
 \beq{-1}
\begin{pmatrix}
 \mathbf{t}_r 
 \\   \\
  \mathbf{t}_{\theta}
  \\  \\
  \mathbf{t}_z  
 \end{pmatrix}
 =
 \begin{pmatrix}
  \mathbf{ { Q}} & \mathbf{R} & \mathbf{P}
  \\ & & \\
   \mathbf{R}^t &  \mathbf{ { T}} & \mathbf{S}
 \\ & & \\
  \mathbf{P}^t &  \mathbf{S }^t &\mathbf{ { M}}
 \end{pmatrix}
 \begin{pmatrix}
 \mathbf{u}_{,\,r}
 \\  \\
  \frac1r(\mathbf{u},_{\,\theta}+\mathbf{K}\,\mathbf{u})
  \\  \\
  \mathbf{u}_{,\,z}
 \end{pmatrix},
 \quad
\begin{pmatrix}
  \mathbf{ { Q}} & \mathbf{R} & \mathbf{P}
  \\ & & \\
   \mathbf{R}^t &  \mathbf{ { T}} & \mathbf{S}
 \\ & & \\
  \mathbf{P}^t &  \mathbf{S }^t &\mathbf{ { M}}
 \end{pmatrix}
=
 \begin{pmatrix}
   (e_r e_r) &  (e_r e_\theta ) &  (e_r e_z )
  \\ & & \\
   &   (e_\theta e_\theta ) &  (e_\theta e_z )
 \\ & & \\
   &    &  (e_z e_z )
 \end{pmatrix}, \nonumber
 \eeq
 }
where $\mathbf{K} = \mathbf{e}_\theta \otimes \mathbf{e}_r
- \mathbf{e}_r\otimes \mathbf{e}_\theta$, and in    notation similar to that of \cite{Lothe76},    the matrix $\left( ab\right) $ has 
components $\left( ab\right) _{jl}=a_{i}C_{ijkl}b_{k}$  for  arbitrary vectors $\mathbf{a}$ and
$\mathbf{b}$.
The explicit form  of the various   matrices is apparent with  the use of  
Voigt's notation
\bal{783}
\mathbf{{Q}}=
\begin{pmatrix}
c_{11}&c_{16}&c_{1\bar 5}
\\
c_{16}&c_{66}&c_{\bar 56}
\\
c_{1\bar 5}&c_{\bar 56}&c_{\bar 5\bar 5}
\end{pmatrix},
\quad 
&
\mathbf{{T}}=
\begin{pmatrix}
c_{\bar 6\bar 6}&c_{2\bar 6}&c_{4\bar 6}
\\
c_{2\bar 6}&c_{22}&c_{24}
\\
c_{4\bar 6}&c_{24}&c_{44}
\end{pmatrix},
\quad
\mathbf{{M}}= 
\begin{pmatrix}
c_{55}&c_{\bar 45}&c_{35}
\\
c_{\bar 45}&c_{\bar 4\bar 4}&c_{3\bar 4}
\\
c_{35}&c_{3\bar 4}&c_{33}
\end{pmatrix},
\nonumber \\
& \\ \nonumber
\mathbf{S}=
\begin{pmatrix}
c_{5\bar 6}&c_{\bar 4\bar 6}&c_{3\bar 6}
\\
c_{25}&c_{2\bar 4}&c_{23}
\\
c_{45}&c_{4\bar 4}&c_{34}
\end{pmatrix},
\quad 
&
\mathbf{P}=
\begin{pmatrix}
c_{15}&c_{1\bar 4}&c_{13}
\\
c_{56}&c_{\bar 46}&c_{36}
\\
c_{5\bar 5}&c_{\bar 4\bar 5}&c_{3\bar 5}
\end{pmatrix},
\quad 
 \mathbf{R}=
\begin{pmatrix}
c_{1\bar 6}&c_{12}&c_{14}
\\
c_{6\bar 6}&c_{26}&c_{46}
\\
c_{\bar 5\bar 6}&c_{2\bar 5}&c_{4\bar 5}
\end{pmatrix} .
\nonumber
\eal
Note that the six  elements in each of the symmetric matrices 
$\mathbf{Q}$, $\mathbf{T}$, $\mathbf{M}$ and the nine in each of 
$\mathbf{S}$, $\mathbf{P}$, $\mathbf{R}$, are independent.  That is, there is one-to-one correspondence between the elements of the six matrices and the 45 independent components of the Cosserat elasticity tensor \eqref{433}. 

\subsubsection{Cylindrically anisotropic materials}

  We consider  materials with no axial dependence whose 
 density and the elasticity tensor  may depend upon    $r$, i.e.
$\rho=\rho(r)$ and $C_{ijkl}=C_{ijkl}(r)$.    
We   seek  solutions  in the form of time-harmonic cylindrical
waves of azimuthal order $n=0,1,2,\ldots $ and axial wavenumber $k_z$, such that the displacement-traction 6-vector is of the form  
 \beq{10}\begin{pmatrix}
\mathbf{u}(r, \theta, z, t)
\\ \\ 
ir \mathbf{t}_r(r, \theta, z, t)
\end{pmatrix}
= \pmb{\eta} (r)\, 
\ee^{i (n\theta  +k_z z - \omega t) } , 
 \eeq
where $\pmb{\eta}$ depends only on the radial coordinate $r$. 
Accordingly, the governing equations of motion  reduce to an ordinary differential equation for this   $6\times 1$ vector  \cite{Shuvalov03nonote}: 
\beq{130}
\frac{\dd\pmb{\eta}  }{\dd r} -    \frac{i}{r}\,\mathbf{G}(r)\pmb{\eta}(r)= 0  ,
 \eeq
with  $6\times 6$ system matrix $\mathbf{G}$,  where 
\bal{020}
i\mathbf{G}(r) & = 
\begin{pmatrix}
- \mathbf{{Q}}^{-1}\widetilde{\mathbf R} 
& - i\mathbf{{Q}}^{-1}
\\ & \\ 
i\big( 
\widetilde{\mathbf T} - \widetilde{\mathbf R}^+ \mathbf{{Q}}^{-1}\widetilde{\mathbf R}
- \rho \omega^2 r^2\big) 
& \widetilde{\mathbf R} ^+ \mathbf{{Q}}^{-1}
\end{pmatrix} 
\nonumber \\ 
& + ik_z r 
\begin{pmatrix}
- \mathbf{{Q}}^{-1} {\mathbf P} 
&  \mathbf{{0}} 
\\ & \\ 
i\big[ \mathbf{P}^t \mathbf{{Q}}^{-1} \widetilde{\mathbf R} -\widetilde{\mathbf S}
-\big( \mathbf{P}^t \mathbf{{Q}}^{-1} \widetilde{\mathbf R} -\widetilde{\mathbf S} \big)^+
+ ik_z r \big( 
\mathbf{P}^t\mathbf{{Q}}^{-1}\mathbf{P} -{\mathbf M}
\big)
\big]
& -  {\mathbf P}^t \mathbf{{Q}}^{-1} 
\end{pmatrix},
\nonumber  \\
& \\ &
\widetilde{\mathbf R} = {\mathbf R}\pmb{\kappa},
\qquad
\widetilde{\mathbf S} = \pmb{\kappa}{\mathbf S},
\qquad
 \widetilde{\mathbf T} = \pmb{\kappa}^+{\mathbf T}\pmb{\kappa} = \widetilde{\mathbf T}^+ ,
 \qquad
 \pmb{\kappa}=\mathbf{K}+ in\mathbf{I} =-\pmb{\kappa}^+ .
 \nonumber
\eal
Note the symmetry $\mathbf{G} = \mathbf{J}\mathbf{G}^+\mathbf{J}$ for  real-valued material constants and $\omega$, $k_z$, where $\mathbf{J}$ has block structure with zero submatrices on the diagonal and
off-diagonal identity matrices.  This hermiticity-like property has  important physical consequences  such as conservation of energy \cite{Norris10}.


\subsubsection{Transformation in cylindrical coordinates}
Let $R=(X_1^2+X_2^2)^{1/2}$, $r=(x_1^2+x_2^2)^{1/2}$, and consider the reverse   deformation 
$R=R(r), X_3=x_3$.  The deformation gradient is then  
\beq{41}
{\pmb F} =  \alpha {\pmb I}_r+  \beta {\pmb I}_\theta + {\pmb I}_z, 
\quad 
\text{with  }\ \
\alpha =  \frac{\dd r}{\dd R}, \quad \beta  = \frac{r}{R}, 
\eeq
where ${\pmb I}_r = {\pmb e}_r\otimes{\pmb e}_r$,  ${\pmb I}_\theta = {\pmb e}_\theta\otimes{\pmb e}_\theta$, 
${\pmb I}_z = {\pmb e}_z\otimes{\pmb e}_z$. 
Taking the (assumed constant) gauge matrix as 
 ${\pmb A}= {\pmb I}$ gives isotropic density, ${\pmb \rho} =  \rho\, {\pmb I}$ and Cosserat elastic stiffness ${\pmb C} $, where according to eq. \eqref{81},
\begin{subequations}
\bal{44}
\rho &= \frac{1}{\alpha \beta}\rho_0 , 
\\
{\pmb C} 
&= \frac{1}{\alpha \beta}
\begin{pmatrix}
	 \alpha^2 c^{(0)}_{11} & \alpha \beta c^{(0)}_{12} & \alpha c^{(0)}_{13} &
	\alpha\beta c^{(0)}_{14} & 	\alpha c^{(0)}_{14} & 	\alpha c^{(0)}_{15} & 	\alpha^2 c^{(0)}_{15} & 	\alpha^2 c^{(0)}_{16} & 	\alpha\beta c^{(0)}_{16} 
	\\ 	&&&&&&&&\\
	  & \beta^2 c^{(0)}_{22} & \beta c^{(0)}_{23} &
	\beta^2 c^{(0)}_{24} & 	\beta c^{(0)}_{24} & 	\beta c^{(0)}_{25} & 	\alpha\beta c^{(0)}_{25} & 	\alpha\beta c^{(0)}_{26} & 	\beta^2 c^{(0)}_{26} 
	\\ 	&&&&&&&&\\
	  &  & c^{(0)}_{33} &
	\beta c^{(0)}_{34} & 	c^{(0)}_{34} & 	c^{(0)}_{35} & 	\alpha c^{(0)}_{35} & 	\alpha c^{(0)}_{36} & 	\beta c^{(0)}_{36} 
	\\ 	&&&&&&&&\\
	       &       &      & \beta^2 c^{(0)}_{44} & \beta c^{(0)}_{4  4}&
\beta c^{(0)}_{45} & 	\alpha\beta c^{(0)}_{45} & 	\alpha\beta c^{(0)}_{46} & \beta^2	c^{(0)}_{46} 
	\\ 	&&&&&&&&\\
         &       &       & &  c^{(0)}_{  4   4}&
c^{(0)}_{45} & 	\alpha c^{(0)}_{45} & 	\alpha c^{(0)}_{46} & 	\beta c^{(0)}_{46} 				
	\\ 	&&&&&&&&\\
         &      &      &       &           & c^{(0)}_{55}   & \alpha c^{(0)}_{5  5} &
		\alpha c^{(0)}_{56} & 	\beta c^{(0)}_{56} 
	\\ 	&&&&&&&&\\
         &       &      &      &      &   &   \alpha^2c^{(0)}_{  5  5}   &
		\alpha^2 c^{(0)}_{56} & 	\alpha\beta c^{(0)}_{56} 
	\\ 	&&&&&&&&\\
	      &   S   &   Y   &    M   &          &       &          & \alpha^2 c^{(0)}_{66}   & \alpha \beta c^{(0)}_{6  6}   
	\\ &&&&&&&&\\
	      &      &      &      &          &        &        &    & \beta^2 c^{(0)}_{  6  6} 
\end{pmatrix}
.  \label{44b}
\eal
\end{subequations}
Equivalently, the six matrices of \eqref{783} that are in one-to-one correspondence with ${\pmb C} $ are 
\bal{787}
\mathbf{Q}&=  \frac{\alpha}{\beta} \mathbf{Q}^{(0)}, 
\quad
\mathbf{T}=  \frac{\beta}{\alpha} \mathbf{T}^{(0)}, 
\quad
\mathbf{M}= \frac {1}{\alpha \beta} \mathbf{M}^{(0)}, 
\nonumber 
\\
\mathbf{S}&= \frac {1}{\alpha}  \mathbf{S}^{(0)}, 
\quad
\mathbf{P}= \frac {1} {\beta}  \mathbf{P}^{(0)}, 
\quad
\mathbf{R}=    \mathbf{R}^{(0)}. 
\eal

These results are consistent with the system equation \eqref{130} in the current coordinates  %
and the analogous equation in the original coordinates, 
\beq{1301}
\frac{\dd \pmb{\eta}}{\dd R}  =    \frac{i}{R}\,\mathbf{G}^{(0)}(R)\, \pmb{\eta}  ,
 \eeq
with 
$\mathbf{G}^{(0)}(R)$ defined in the same way as \eqref{020}  using 
$\mathbf{Q}^{(0)}$, $\ldots$, $\mathbf{R}^{(0)}$.  Note that the transformed version of \eqref{1301} may be  
obtained directly by multiplication with the factor $\dd R/\dd r$,  
\beq{1302}
\frac{\dd \pmb{\eta}}{\dd r}  =    \frac{i}{R}\frac{\dd R}{\dd r}\,\mathbf{G}^{(0)}(R)\, \pmb{\eta} .
 \eeq
Comparison of eqs. \eqref{130} and \eqref{1302}
implies that the transformed system matrix is 
\beq{1303}
 \mathbf{G}(r) =    \frac{r}{R}\frac{\dd R}{\dd r}\,\mathbf{G}^{(0)}(R) . 
 \eeq
This relation, combined with the block structure of $\mathbf{G}$ in \eqref{020} and the analogous form for  $\mathbf{G}^{(0)}$,   implies the matrix relations \eqref{787}. 

\subsection{Example: a cylindrically orthotropic material}\label{5.2}

\subsubsection{Transformed elastic moduli}

The initial material is assumed to be cylindrically orthotropic, with stiffness in the usual Voigt notation ($\{1,2,3,4,5,6\} \leftrightarrow  \{11,22,33,23,31,12\}$) given by 
\beq{43}
{\pmb C}^{(0)} =\begin{pmatrix}
	c_{11}^{(0)} &c_{12}^{(0)} &c_{13}^{(0)} &0 &0 &0 \\
	c_{12}^{(0)} &c_{22}^{(0)} &c_{23}^{(0)} &0 &0 &0 \\
	c_{13}^{(0)} &c_{23}^{(0)} &c_{33}^{(0)} &0 &0 &0 \\
	0 &0 &0 &c_{44}^{(0)} &0 &0\\
  0 &0 &0&0 &c_{55}^{(0)}  &0\\
 0 &0 &0&0&0 &c_{66}^{(0)}  
\end{pmatrix}.  
\eeq
Hence, using \eqref{44b}, 
the transformed elastic stiffness tensor becomes in the 9-index Cosserat notation, 
\beq{447}
{\pmb C} 
= \frac {1}{\alpha \beta}
\begin{pmatrix}
	 \alpha^2 c^{(0)}_{11} & \alpha \beta c^{(0)}_{12} & \alpha c^{(0)}_{13} &0      &0          &0                  &0 &0 &0 \\
	  & \beta^2 c^{(0)}_{22} & \beta c^{(0)}_{23} &0      &0          & 0                 &0 &0 &0 \\
	  &  & c^{(0)}_{33} &0      &0          &0                  &0 &0 &0\\
	       &       &      & \beta^2 c^{(0)}_{44} & \beta c^{(0)}_{4  4}&0                  &0&0&0\\
         &       &       & &  c^{(0)}_{  4   4}&0           &0     &0           &0\\
         &      &      &       &           & c^{(0)}_{55}   & \alpha c^{(0)}_{5  5} &0  &0  \\
         &       &      &      &      &   &   \alpha^2c^{(0)}_{  5  5}   &0 &0  \\
	      &   S   &   Y   &    M   &          &       &          & \alpha^2 c^{(0)}_{66}   & \alpha \beta c^{(0)}_{6  6}   \\
	      &      &      &      &          &        &        &    & \beta^2 c^{(0)}_{  6  6} 
\end{pmatrix}
.
\eeq

\subsubsection{SH motion}

The equation  of motion in the undeformed coordinates for  ${\pmb U} = (0,0,U(r,\theta,t))$, 
\beq{46}
\frac {1}{R} \big(R c^{(0)}_{55} U_{,R}\big)_{,R}+ \frac{1}{R^2}\big( c^{(0)}_{44} U_{,\theta}\big)_{,\theta} -\rho_0\ddot{U}=0, 
\eeq
transforms to the following 
equation for  ${\pmb u} = (0,0,u(r,\theta ,t))$,  
\beq{451}
\frac 1 r (rc_{\bar 5 \bar 5} u_{,r})_{,r}+ \frac 1 {r^2} (c_{44}u_{,\theta})_{,\theta} -\rho\ddot{u}=0, 
\eeq
where
\beq{47}
\rho =  \rho_0 \frac{R}{r} \frac{\dd R}{\dd r} , 
\quad
c_{44} =  c^{(0)}_{44}  \frac{r}{R} \frac{\dd R}{\dd r}, 
\quad
c_{\bar 5 \bar 5} =  c^{(0)}_{55}  \frac{R}{r}\frac{\dd R}{\dd r}.  
\eeq
This is an example of the general result in \S \ref{-0-0} for SH motion in a plane of symmetry .

\subsubsection{In-plane  motion}
The displacement is assumed to have the form  ${\pmb u} = (u_r(r,\theta ,t),u_\theta (r,\theta ,t) ,0)$.   The density is again isotropic given by eq. \eqref{47}$_1$ and the relevant moduli are, from eq. \eqref{44b}, 
\begin{subequations}\label{48}
\bal{48a}
c_{11}&=   c^{(0)}_{11} \frac{R}{r}\frac{\dd R}{\dd r},\quad 
c_{22} =  c^{(0)}_{22} \frac{r}{R}\frac{\dd R}{\dd r},\quad 
c_{12} =    c^{(0)}_{12},\\
c_{66}&=  c^{(0)}_{66} \frac{R}{r} \frac{\dd r}{\dd R},\quad 
c_{\bar 6\bar 6}=  c^{(0)}_{66} \frac{r}{R}\frac{\dd R}{\dd r},\quad 
c_{6\bar 6}=    c^{(0)}_{66}. \label{48b}
\eal
\end{subequations}
For isotropic initial stiffness tensor, these expressions  agree with the Cosserat elastic moduli found by \cite{Brun09}, where only the particular transformation of \cite{Greenleaf03,Pendry06} was considered, i.e., $R=r_1(r-r_0)/(r_1-r_0)$ for $r\in (r_0,r_1]$.  

\subsection{A symmetric approximation for $k_z = 0$ }\label{sec6}
The example in \S \ref{5.2} achieves cloaking of in-plane elastic wave motion using elastic moduli 
of Cosserat form, \eqref{48}, generalizing the findings of \cite{Brun09}.  The existence of three distinct in-plane shear moduli, 
$c_{1212}$, $c_{2121}$, $c_{1221}=c_{2112}$, is a property of the Cosserat model, but one that is difficult if not impossible to realize in practice.   We now show that there is a preferred approximation with only a single shear modulus.

Consider the transformed system matrix $\mathbf{G} $  with $k_z = 0$ for the initial cylindrically orthotropic material \eqref{43}, which  follows from \eqref{1303}   as  
\beq{5-5}
i \mathbf{G} =  \frac{\beta}{\alpha} 
\begin{pmatrix}
 - \begin{pmatrix}
0  &    {c^{(0)}_{11}}^{-1} c^{(0)}_{12}  &0
\\
1&0&0 
\\
0&0&0
\end{pmatrix}
\pmb{\kappa}
 & -i
\begin{pmatrix}
 {c^{(0)}_{11}}^{-1} &0&0
\\
0&  {c^{(0)}_{66}}^{-1} &0
\\
0& 0&  {c^{(0)}_{55}}^{-1} 
\end{pmatrix}
\\
i \pmb{\kappa}^+
\begin{pmatrix}
0 &0&0
\\
0&  c^{(0)}_{22} -{c^{(0)}_{11}}^{-1}{c^{(0)}_{12}}^{2} &0
\\
0& 0&  {c^{(0)}_{44}}
\end{pmatrix}
\pmb{\kappa}
- i \omega^2 R^2 \rho_0 {\pmb I}
&
- \pmb{\kappa}
\begin{pmatrix}
0  &    1   &0
\\
{c^{(0)}_{11}}^{-1} c^{(0)}_{12} &0&0 
\\
0&0&0
\end{pmatrix}
\end{pmatrix}
.
\eeq
Let $\overline C_{ijkl}$ be a set of moduli in the transformed coordinates, 
corresponding to a normal elastic solid (symmetric stress) of  cylindrically orthotropic symmetry, with 
moduli (in the standard Voigt notation), 
\beq{4343}
 \overline {\pmb C}  =\begin{pmatrix}
	\overline c_{11}  & \overline c_{12}  &\overline c_{13}  &0 &0 &0 \\
	\overline c_{12}  &\overline c_{22}  &\overline c_{23}  &0 &0 &0 \\
	\overline c_{13}  &\overline c_{23}  &\overline c_{33}  &0 &0 &0 \\
	0 &0 &0 &\overline c_{44}  &0 &0\\
  0 &0 &0&0 &\overline c_{55}   &0\\
 0 &0 &0&0&0 &\overline c_{66}   
\end{pmatrix}   .
\eeq
Assuming the density is isotropic and equal to the transformed density of eq. \eqref{44}, 
the  system matrix  associated with the stiffness $\overline C_{ijkl}$ becomes 
\beq{5-6}
i \overline{\mathbf{G}} =  
\begin{pmatrix}
 - \begin{pmatrix}
0  &    { {\overline c}_{11}}^{-1} {\overline c}_{12}  &0
\\
1&0&0 
\\
0&0&0
\end{pmatrix}
\pmb{\kappa}
 & -i
\begin{pmatrix}
 {{\overline c}_{11}}^{-1} &0&0
\\
0&  {{\overline c}_{66}}^{-1} &0
\\
0& 0&  {{\overline c}_{55}}^{-1} 
\end{pmatrix}
\\
i \pmb{\kappa}^+
\begin{pmatrix}
0 &0&0
\\
0&  {\overline c}_{22} -{{\overline c}_{11}}^{-1}{{\overline c}_{12}}^{2} &0
\\
0& 0&  {{\overline c}_{44}}
\end{pmatrix}
\pmb{\kappa}
- i \omega^2 r^2 \rho {\pmb I}
&
- \pmb{\kappa}
\begin{pmatrix}
0  &    1   &0
\\
{{\overline c}_{11}}^{-1} {\overline c}_{12} &0&0 
\\
0&0&0
\end{pmatrix}
\end{pmatrix}
.
\eeq
Comparison of \eqref{5-5} and \eqref{5-6} suggests the identification
\bse{2-4}
\bal{2-4a}
{\overline c}_{11} &= \frac\alpha\beta c^{(0)}_{11} , 
\quad
{\overline c}_{22} = \frac\beta\alpha c^{(0)}_{22}, 
\quad
{\overline c}_{12} =   c^{(0)}_{12} , 
\\
{\overline c}_{44} &= \frac\beta\alpha c^{(0)}_{44} , 
\quad
{\overline c}_{55} = \frac\beta\alpha c^{(0)}_{55}, 
\quad
{\overline c}_{66} = \frac\alpha\beta c^{(0)}_{66} . 
\eal
\ese
This set of moduli has the property that they correspond to a normal elastic material, as compared with the  Cosserat material required for the exact solution. Comparing the  latter, given by \eqref{447}, with 
the proposed moduli \eqref{2-4} shows that 
\beq{2-46} 
{\overline c}_{11} =  c_{11} , 
\ \
{\overline c}_{22} =   c_{22}  ,  
\ \
{\overline c}_{12} =     c_{12} , 
\ \
{\overline c}_{44} =   c_{44}  , 
\ \
{\overline c}_{55} =   c_{\bar 5\bar 5} , 
\ \
{\overline c}_{66} = c_{66}  . 
\eeq
Let $\overline{\pmb{\eta}} (r) $ be the solution for the approximate but symmetric moduli \eqref{4343}, 
\beq{1305}
\frac{\dd \overline{\pmb{\eta}} }{\dd r}      -
\frac{i}{r}\,\overline{\mathbf{G}}(r)\overline{\pmb{\eta}}(r) 
 =0 . 
 \eeq
 
 It is evident that the exact and approximate systems \eqref{130} and \eqref{1305}  have identical SH solutions.   
 The approximate solution defined by the moduli $\overline C_{ijkl}$   possesses  a further  interesting property related to in-plane wave motion.  Thus,  
{the difference between the system matrices of the exact transformed medium and that of the approximate material follows as }
\bal{5-8}
 \frac ir {\mathbf{G}} - \frac ir  \overline{\mathbf{G}} &=  f(r) \mathbf{D} \equiv \mathbf{\Delta }, 
 \\
 \quad \text{where }\   
 f( r) =\frac{1}{R}\frac{\mathrm{d}R}{\mathrm{d}r}-\frac{1}{r},
 \quad 
 \mathbf{D} &= 
\begin{pmatrix}
 \mathbf{D}_1 & \mathbf{0}
\\
 \mathbf{0} & -\mathbf{D}_1^+
\end{pmatrix},
\ \
\mathbf{D}_1 = 
\begin{pmatrix}
 0  &    0  &0  
\\
-in&1&0 
\\
0&0&0 
\end{pmatrix}
.
\eal
The difference is independent of the material properties, 
 a function  of only the transformation function through $f(r)$ and 
 of the azimuthal index $n$ through $\mathbf{D}$ which is rank two and has  the property 
 $\mathbf{D}^k = \mathbf{D}$   or $\mathbf{D}^2$ for any odd or even $k$.    This suggests that the choice \eqref{2-4} of the approximate material parameters is in some sense preferred over others.   Some consequences are discussed next. 

The matricant solutions $\mathbf{M}(r,r_1)$ and $\overline{\mathbf{M}}(r,r_1)$ of,
respectively,  the exact and approximate systems \eqref{130} and \eqref{1305}, 
are defined \cite{Pease}  such that $\mathbf{\eta} =\mathbf{ Mc}$  and  $\overline{\mathbf{\eta }}=\overline{\mathbf{\mathbf{M}}}\mathbf{c}$  where  $\mathbf{c}$  is the arbitrary 
initial data vector at $r=r_1$.  The matricants therefore satisfy differential equations similar to the state vectors with initial conditions $\mathbf{M}(r_1,r_1) = \mathbf{I}$, $\overline{\mathbf{M}}(r_1,r_1)= \mathbf{I}$.
According to \cite[\S 3 of Ch.VII]{Pease}, the
matricant  \eqref{130} and \eqref{1305} on an interval $\left[ r_{1},r_{2}\right] $ are
related as 
$\mathbf{\mathbf{M}=\overline{\mathbf{M}}P}$ through the matrix 
$\mathbf{P}$ satisfying $\mathbf{P}^{\prime }= \overline{\mathbf{M}} ^{-1} \mathbf{\Delta} \overline{\mathbf{M}} \mathbf{P}$.   Noting that 
$\mathbf{R}^{\prime }=  \mathbf{\Delta}  \mathbf{R}$  with initial condition $\mathbf{R}(r_1,r_1) = \mathbf{I}$ has a simple explicit solution
\beq{3=11}
\mathbf{R}^{\pm 1} =\mathbf{I}+\mathrm{diag}\bigg( 
\big[ \big( \frac{rR_1}{R r_1}\big)^{\pm 1} -1\big] \mathbf{D}_{1}
,\,  
\big[ \big( \frac{rR_1}{R r_1}\big)^{\mp 1} -1\big] \mathbf{D}_{1}^{+} \bigg) , 
\eeq
it follows that  
\beq{3=7-}
\mathbf{M}=\overline{\mathbf{M}}\mathbf{R} \mathbf{T} \quad\text{where   }\ 
\mathbf{T}' = \mathbf{\Delta}_1 \mathbf{T}
, \quad \mathbf{T}(r_1,r_1) = \mathbf{I}, 
\eeq
and   $\mathbf{\Delta}_1 = (\overline{\mathbf{M}}\mathbf{R} )^{-1} ( 
\mathbf{\Delta}\overline{\mathbf{M}} -\overline{\mathbf{M}} \mathbf{\Delta}) \mathbf{R}$. 
The latter vanishes  at $r=r_1$, which suggests that the  convergent Peano series    $\mathbf{T}
= \mathbf{T}_0 + \mathbf{T}_1 + \mathbf{T}_2 +\ldots$ with $\mathbf{T}_0
= \mathbf{I}$, $\mathbf{T}_j' = \mathbf{\Delta}_1 \mathbf{T}_{j-1}$, $j=1,2, \ldots$,   forms a natural and regular perturbation solution. 

Note that in cylindrical cloaks the original domain   $\left[ R_{1}< r_1,R_2 = r_{2}\right]$ is mapped to the smaller one $\left[ r_{1},r_{2}\right] $ so that the  elements of $\mathbf{R}$ in 
\eqref{3=11} are sign definite.


\section{Discussion}\label{sec7}

A complete transformation theory has been  developed for elasticity. 
The  material properties after transformation of the elastodynamic equations  are  given by eqs. \eqref{2-1} through \eqref{-51}.   The constitutive parameters  depend on both the transformation and gauge matrices, ${\pmb F}$ and  ${\pmb A}$, and do not necessarily have symmetric stress. 
  It was shown  in \S \ref{sec3} that   {\it{a priori}}  requiring  stress to be 
    symmetric  implies that the material must be of Willis form \eqref{-272}, with ${\pmb A}={\pmb F}$ as Milton et al. \cite{Milton06} found.    The emphasis here has been on exploring the consequences of relaxing the constraint of symmetric stress.  
There are several reasons for doing so.  First  is the fact that the transformation of the acoustic equation in its simplest form, i.e. by identifying an inertial tensor ${\pmb \rho} = J^{-1} {\pmb F}{\pmb F}^t$ from the differential identity 
$
 \Div\Grad f  \rightarrow   J\div J^{-1} {\pmb F}{\pmb F}^t\grad f$, does not follow from the transformed Willis material, even though the inertial
 fluid has symmetric stress.  Other types of transformed acoustic fluids are possible (e.g. pentamode materials), again with symmetric stress and  not contained within the framework of  eqs. \eqref{-272}.  A second and more practical reason for
 considering the general material as defined by eqs. \eqref{2-1} through \eqref{-51} with  ${\pmb F}$ and  ${\pmb A}$ distinct is to broaden the class of  materials  available for design of elastodynamic cloaks.   
 
Allowing ${\pmb A}$ to be  independent of ${\pmb F}$ leads to constitutive models that differ markedly from the Willis material model.  In this paper  we have emphasized solutions corresponding to time-independent material parameters  obtained  when  ${\pmb A}$ is assumed constant   
 (see  eqs. \eqref{8},  \eqref{2-1} and \eqref{-51a}),
\beq{-4-}
{\pmb \rho} =\rho_0 \, J^{-1}  {\pmb A}{\pmb A}^t,
\quad
 C_{ijkl}
=J^{-1} C^{(0)}_{ IJKL}\,  F_{iI}A_{jJ} F_{kK}A_{lL} .
\eeq
   These transformed quantities correspond to a material with anisotropic density tensor  and  stress-strain relation of Cosserat type (non-symmetric stress).  
  Setting ${\pmb A}= {\pmb I}$ ensures that the transformed density is isotropic (see eq. \eqref{81})
	\beq{-81}
 \rho =  J^{-1} \, \rho_0 ,
\quad
 C_{ijkl}^\text{eff} 
=J^{-1} C^{(0)}_{ IjKl}\,  F_{iI} F_{kK} .  
\eeq
However, there is no general choice of the  gauge matrix ${\pmb A}$ that will make the stress symmetric for a given transformation.  This feature distinguishes the elastic transformation problem from the acoustic case, for which  it is always possible to achieve symmetric, even isotropic (hydrostatic), stress.   If the transformation is homogeneous, corresponding to constant ${\pmb F}$, it is possible to make the elastic stress symmetric, although at the price of anisotropic inertia (see eq. \eqref{813}).    


Materials displaying non-symmetric stress of the type necessary to achieve elastodynamic cloaking   while  difficult to envisage,  are not ruled out.    Effective moduli with the major symmetry 
$C_{ijkl} = C_{klij}$ that do not display the minor symmetry  $C_{ijkl} = C_{jikl}$ are found in the theory of incremental motion superimposed on finite deformation \cite{Ogden07a}.  The similarity with the transformation problem is intriguing:  small-on-large motion in the presence of finite prestress corresponds to a transformation of an actual material via a deformation.   The deformation in the small-on-large theory is however quite distinct from  ${\pmb F}$ in the present context.  
The formal equivalence of  the constitutive parameters \eqref{-81} with  the  density and moduli for incremental motion after finite  prestress offers the possibility for achieving Cosserat elasticity of the desired form.  The crucial quantity is the finite (hyperelastic) strain energy function of the material, which after prestress should have the desired Cosserat incremental moduli.  Future work will examine this connection and the types of strain energy functions required. 

Another approach is to seek  materials with normal elastic behavior that approximate, in some sense, the Cosserat material.  Preliminary work in this direction has been considered here. 
The general theory has been applied to the case of cylindrical anisotropy for  arbitrary radial transformation $R \rightarrow r$.   The equations of motion for the transformed Cosserat material have been expressed in Stroh format,  eq. \eqref{130}, suitable for numerical implementation.      The material required for cloaking of in-plane elastic waves is of Cosserat type with isotropic density.  A normal elastic material with density \eqref{44} and elastic moduli defined by eqs.  \eqref{4343} and \eqref{2-4}   appears to provide a  natural approximation.  The properties of this type of approximate material is the subject for planned further study, analytical and numerical.  


%

\end{document}